\documentclass[aps,prl,twocolumn,showpacs,preprintnumbers,amsmath,amssymb,superscriptaddress]{revtex4}
\usepackage{mathptm}  
\usepackage{dcolumn}                    
\usepackage{bm}                        
\usepackage{graphicx}
\usepackage{times}
\usepackage{color}

\newcommand{\vs}{\boldsymbol{\sigma}} 

\newcommand{\e}[1]{\mathrm{e}^{#1}}

\newcommand{\addAB}[1]{\textcolor{black}{#1}}

\newcommand{\eg}{\textit{e.g. }}
\newcommand{\etal}{\textit{et al.}}
\def\i{\mathrm{i}}

\begin{document}
\title{
Spin-Transfer and Exchange Torques in Ferromagnetic Superconductors
}

 \author{Jacob Linder}
\affiliation{Department of Physics, Norwegian University of Science and Technology, N-7491 Trondheim, Norway}
\author{Arne Brataas}
\affiliation{Department of Physics, Norwegian University of Science and Technology, N-7491 Trondheim, Norway}
\author{Zahra Shomali}
\affiliation{Department of Physics, Institute for Advanced Studies
in Basic Sciences (IASBS), Zanjan 45137-66731, Iran}
\author{Malek Zareyan}
\affiliation{Department of Physics, Institute for Advanced Studies
in Basic Sciences (IASBS), Zanjan 45137-66731, Iran}

\date{Received \today}
\begin{abstract}
We consider how superconducting correlations influence spin-transfer torques in ferromagnetic superconductors.
It is demonstrated that there is a novel torque arising from particle-hole interference that depends on the U(1) phase associated with the superconducting order parameter. 
We also show that there is an equilibrium exchange
torque between two ferromagnetic superconductors in contact via a normal metal mediated by Andreev states. The latter equilibrium magnetic torque is also sensitive to spin-resolved phase differences in the superconducting order parameters as well as to an externally applied phase difference.
\end{abstract}

\maketitle

\textit{Introduction.} The synthesis of materials with magnetic and superconducting order
offers interesting possibilities. The combination of 
spin-filtering in ferromagnets with
dissipationless currents
in superconductors
is of a fundamental interest and offers routes towards
novel types of controlled charge and spin flow. A considerable activity in the field focus on the range of the superconducting proximity effects, demonstrating that singlet Cooper pairs can be converted to triplet Cooper pairs in inhomogenous \cite{BergeretRev:05, KeizerNat:06,buzdin_rmp_05,robinson_science_10} or time-dependent \cite{HouzetPRL:08, SkadsemPRB:11, holmquist} magnetic textures. 

In ferromagnets, spin-transfer torques  attract a large interest 
 since they involve the coupling between itinerant carriers and collective magnetic order parameters and can be useful in 
magnetic random access memories and oscillator circuits  \cite{stt,stt2}. Spin-transfer torque\addAB{s result from} 
the transfer of spin angular momentum from the (spin) current to the magnetization\addAB{. }
While many aspects of how $s$-wave superconductivity affects 
spin-transport and spin-transfer torques are known \cite{SpinTransSuper}, there are no predictions of how 
spin-transfer torques are manifested in ferromagnetic superconductors. In these systems, we show that the 
spin-transfer torques depend on the phase of the superconducting pairing correlations. This can be utilized as an additional way of controlling and detecting spin-transport and magnetization dynamics.

In this Letter, we compute magnetic torques in 
ferromagnetic superconductors,
both in equilibrium and out-of-equilibrium. Out-of-equilibrium, a spin-polarized current with a polarization that is non-collinear to the magnetization and injected from a
a normal metal  (N) towards
a ferromagnetic superconductor (FS) generates a novel 
torque on the magnetization of the FS due to particle-hole interference, which strongly depends on the phase of the spin-triplet superconducting order parameter. We further demonstrate there is 
an equilibrium \addAB{magnetic} 
torque in a 
FS/N/FS
Josephson contact mediated by Andreev states. Here, the formation of spin-triplet electron-hole Andreev bound states with a non-collinear spin polarization plays the essential role. We demonstrate the penetration of this equilibrium torque into the FS, and its sensitivity to an external applied phase difference as well as the spin-resolved phase of each FS.

\textit{Theory.} 
To model the coexistence of bulk superconductivity and ferromagnetism, as experimentally verified in UGe$_2$ \cite{saxena_nature_00}, URhGe \cite{aoki_nature_01}, and UCoGe \cite{huy_prl_07}, we consider equal spin-pairing triplet superconductivity. Then, 
Cooper pairs are not broken by 
Zeeman-field\addAB{s} smaller than 
\addAB{70} meV in 
UGe$_2$ \cite{saxena_nature_00}. The variation of the equilibrium exchange interaction between two
ferromagnets with the relative angles of the magnetizations is a Fermi
surface property \cite{bruno_prb_95}. Similarly, the \addAB{out-of}
-equilibrium spin-transfer torque is governed by states near the Fermi level.

Let us first demonstrate that the out-of-equilibrium spin-transfer in ferromagnetic superconductors is qualitatively different than in conventional ferromagnets. Conventionally,  the spin-transfer torque exerted on the magnetic order parameter equals the loss of transverse spin current in the ferromagnet. This absorption takes place over a small distance from the interface region, typically of the order a few Fermi wavelengths in strong ferromagnets. In contrast, in ferromagnetic superconductors, we find that the spin-transfer torque does not equal the loss of quasiparticle spin-current. The underlying reason for this can be understood by inspecting the spin continuity equation. We start by defining the spin-density $\boldsymbol{S}$ and the Hamiltonian $H$:
\begin{align}
\boldsymbol{S} = \frac{1}{2} \psi^\dag \begin{pmatrix}
\vs & 0 \\
0 & -\vs^* \\
\end{pmatrix}\psi,\;\;
H = \begin{pmatrix}
H_0& \Delta \\
\Delta^* & -H_0^* \\
\end{pmatrix},
\label{Ham}
\end{align}
where $\hbar=1$ and $H_0=-\nabla^2/(2m) - \mu - \boldsymbol{h}\cdot\vs$,
$\Delta = \text{diag}(\Delta_\uparrow,\Delta_\downarrow)$. Here,
$\boldsymbol{h}$ is the exchange field, $\vs$ is a vector of Pauli matrices, and $\Delta_\sigma$, $\sigma=\uparrow,\downarrow$ are the superconducting order parameters for majority and minority spin carriers. The Hamiltonian 
(\ref{Ham}) \addAB{determines} 
the rate of change of the spin density\addAB{:}
\begin{align}\label{eq:spineq}
\partial_t \boldsymbol{S} + \partial_i\boldsymbol{J}_S^i = \boldsymbol{\mathcal{S}}_\text{super} + \boldsymbol{\tau}_\text{STT},
\end{align}
where 
\begin{subequations}
\begin{align}
\boldsymbol{J}_S^i &= \frac{1}{2m} \text{Im} \{ \psi_1^\dag \vs \partial_i \psi_1 + \psi_2^\dag \vs^*\partial_i \psi_2 \},\label{eq:termsa}\\
\boldsymbol{\mathcal{S}}_\text{super} &= -\text{Im}\{ \psi_2^\dag \Delta^* \vs \psi_1 - \psi_1^\dag \Delta \vs^* \psi_2\},\label{eq:termsb}\\
\boldsymbol{\tau}_\text{STT} &=  \psi_1^\dag [\vs\times \mathbf{h}] \psi_1 - \psi_2^\dag[\vs^*\times\mathbf{h}]\psi_2\label{eq:termsc}.
\end{align}
\end{subequations} 
and $\psi_1$ and $\psi_2$ are electron- and hole-like $2\times 1$ spinors constituting the total wavefunction, i.e. $\psi = (\psi_1,\psi_2)^T$. 

The \addAB{rate of change of the spin-density} 
(\ref{eq:spineq}) \addAB{consists of} 
the quasiparticle spin-current tensor \addAB{$\boldsymbol{J}_S$} [superscript $i$ indicating its spatial components in Eq. (\ref{eq:spineq})], 
the spin supercurrent carried by the condensate \addAB{$\boldsymbol{\mathcal{S}}_\text{super}$}, 
\addAB{and} the \addAB{spin-transfer} torque exerted on the ferromagnetic order parameter \addAB{$\boldsymbol{\tau}_\text{STT}$}. 
The \addAB{spin-transfer} torque of Eq. (\ref{eq:termsc}) has a simple interpretation in the case \addAB{of stationary transport in }
a normal
metal-ferromagnet system when it represents the loss of the transverse
component of the spin current\addAB{,}
$\partial_i\boldsymbol{J}_S^i = \boldsymbol{\tau}_\text{STT}.$ 
\addAB{Then, }
the total torque 
is $\int \boldsymbol{\tau}_\text{STT} = \boldsymbol{J}_S(\text{F}) - \boldsymbol{J}_S(\text{N})$ where $\boldsymbol{J}_S(\text{N})$ is the spin
current at the N-F interface and $\boldsymbol{J}_S(\text{F})$ is the spin current deep inside the
ferromagnet. In metallic ferromagnets in good contact
with normal metals, the incoherence between the spin-up and spin-down states within
the ferromagnet implies that the transverse components of $\boldsymbol{J}_S(\text{F})$ vanish at length scales that are
larger than the transverse decoherence length. Thus, $\int \boldsymbol{\tau}_\text{STT} =  \boldsymbol{m}\times[\boldsymbol{m}\times\boldsymbol{J}_S(\text{N})]$ 
which is the established consensus 
\cite{stt2}.

Since $\psi_1$ and $\psi_2$ contain 
contribution\addAB{s} from 
electron-
and hole-like quasiparticles,
Eq.~\ref{eq:termsc} shows that the torque is directly modified by superconducting correlations\addAB{.} 
\addAB{In turn, these correlations are controlled by the}
coherence factors \addAB{that} depend explicitly on the superconducting U(1) phases associated with each of the order parameters $\Delta_\sigma$ in $p$-wave ferromagnetic superconductors. This implies that the spin-transfer torque is sensitive to the superconducting phase, in contrast to \eg the charge conductance which is insensitive to the U(1) phase. The origin of this effect 
\addAB{is that the torque acquires} 
contributions from interference-terms \addAB{of the propagation of electron- and hole-like excitations.}
Since these excitations 
have different U(1) superconducting phases due to the spin-resolved condensate, the torque will depend explicitly on the internal phase difference between the two spin condensates. We 
explicitly 
verify this statement \addAB{below.}
\addAB{Since a part of the spin-current is carried by the condensate via}
$\boldsymbol{\mathcal{S}}_\text{super}$, the loss of \addAB{the} quasiparticle spin-current is not fully compensated by the torque $\boldsymbol{\tau}_\text{STT}$ exerted on the ferromagnetic order parameter. 

\begin{figure}[t!]
\centering
\resizebox{0.5\textwidth}{!}{
\includegraphics{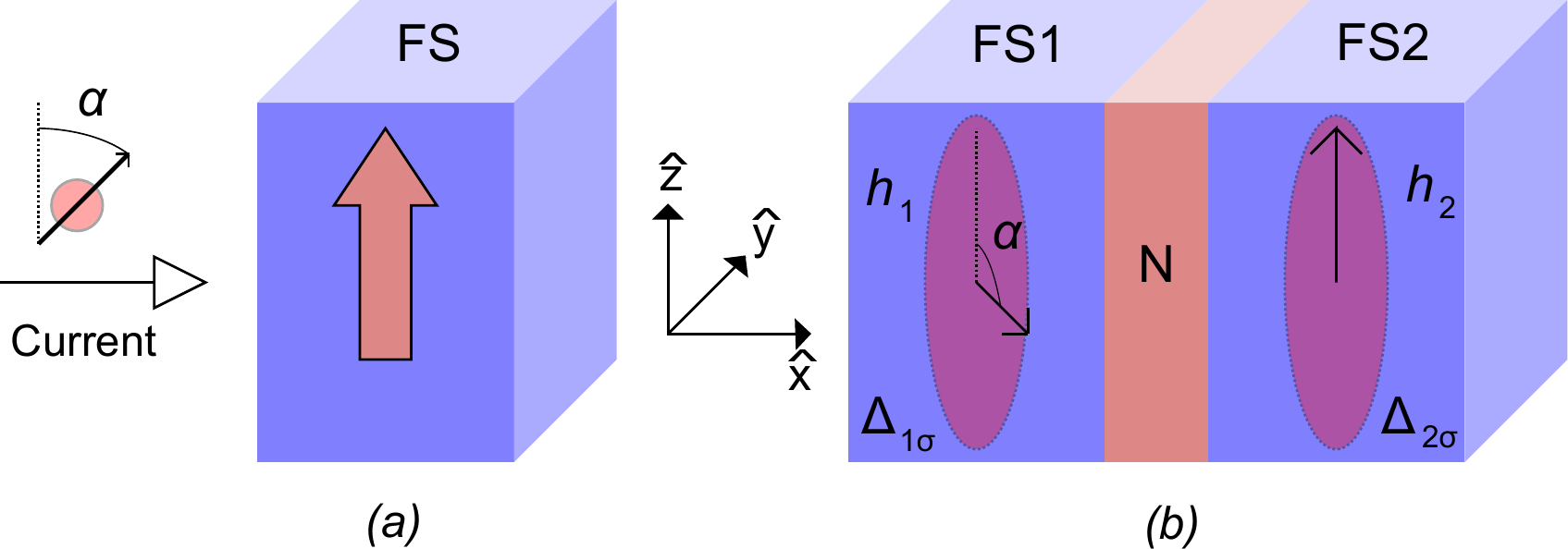}}
\caption{(Color online) (a) Spin-injection into a ferromagnetic superconductor (FS). (b) Exchange torque setup for two FS separated by a normal region. }
\label{model} 
\end{figure}

To explicitly compute the 
spin-transfer torque, we consider a two-dimensional system in the  $x$-$y$ plane and assume there is a normal metal to the left ($x<0$) and a ferromagnetic superconductor to the right ($x>0$). To model spin-injection into the ferromagnetic superconductor, consider an incident particle in the normal metal at the Fermi energy with a magnetic moment at an an angle $\alpha$ with respect to the $z$-axis;
a superposition of spin up and down states along the $z$ direction. Taking into account both normal and Andreev reflection as well as the transverse wave vector
$k_y$, the total wavefunction in the normal metal region in spin-Nambu space is:
$\psi_\text{inc} = [c,s,0,0] \e{\i k_xx} + r_\uparrow[1,0,0,0]\e{-\i k_xx} + r_\downarrow[0,1,0,0]\e{-\i k_xx} + r_A^\uparrow[0,0,1,0]\e{\i k_xx} + r_A^\downarrow[0,0,0,1]\e{\i k_xx},$
where $c=\cos(\alpha/2)$ and $s=\sin(\alpha/2)$.
The longitudinal wavevector is $k_x = \sqrt{k_{FN}^2-k_y^2}$
with $k_{FN}$ being the Fermi wavevector in the normal region. In the ferromagnetic superconductor, the wavefunction is: $\psi_\text{trans} = t_e^\uparrow [u_\uparrow,0,v_\uparrow\e{-\i\gamma_+^\uparrow-\i\theta_\uparrow},0]\e{\i q_e^\uparrow x} +
t_e^\downarrow[0,u_\downarrow,0,v_\downarrow\e{-\i\gamma_+^\downarrow-\i\theta_\downarrow}]\e{\i q_e^\downarrow x}+ t_h^\uparrow[
v_\uparrow\e{\i\gamma_-^\uparrow+\i\theta_\uparrow},0,u_\uparrow,0]\e{-\i q_h^\uparrow x}
+ t_h^\downarrow[0,v_\downarrow\e{\i\gamma_-^\downarrow+\i\theta_\downarrow},0,u_\downarrow]\e{-\i q_h^\downarrow x}$.
In this expression, we have as an illustration assumed chiral $p$-wave superconducting gaps similar to the A2-phase in liquid $^3$He \cite{leggett_rmp_75}, $\Delta_\sigma = \Delta_{\sigma,0}(k_x+\i k_y)/k_F$
and it is straightforward to consider other scenarios.
Furthermore, we have defined
$q_e^\sigma = \sqrt{k_{FS}^2 + 2m(\sigma h + \i|\Delta_\sigma|)-k_y^2},
q_h^\sigma = \sqrt{k_{FS}^2 + 2m(\sigma h - \i|\Delta_\sigma|)-k_y^2}$ 
and
$\e{\delta\i\gamma_\beta^\sigma} = \frac{(\beta k_x + \i \delta k_y)}{\sqrt{k_{FS}^2+2m\sigma h}},\; \beta,\delta=\pm1, \sigma=\uparrow,\downarrow,$ while $k_{FS}$ is the normal-state Fermi wavevector in the superconducting region. The transmission and reflection coefficients, which allows computation of the transport properties, are obtained by applying continuity of the wave functions and currents at the interface.

Experimentally, the relations $\mu\gg h\gg|\Delta_\sigma|$ hold: the superconducting transition occurs only deep within the ferromagnetic phase. To reduce the complexity of the analytical results, we assume that $k_{FS} \gg k_{FN}$; an assumption that has no qualitative effects on our main findings. With these assumptions, the torque in Eq. (\ref{eq:termsc}) is:
\begin{widetext}
\begin{align}\label{eq:tau}
\Big(\boldsymbol{\tau}_\text{STT}\Big)_x &= 8hk_x^2cs\e{-x/\xi_S} \text{Im} \Big\{ \mathcal{A}_1\e{-2\i hx/v_F} (u_\uparrow^*u_\downarrow + v_\uparrow^*v_\downarrow\e{\i(\gamma_+^\uparrow-\gamma_+^\downarrow)+\i\Delta\theta}) + \mathcal{A}_2\e{2\i hx/v_F} (u_\uparrow^*u_\downarrow + v_\uparrow^*v_\downarrow\e{\i(\gamma_-^\downarrow-\gamma_-^\uparrow)+\i\Delta\theta})\notag\\
&\mathcal{A}_3\e{\i(2k_{FS}^2-k_y^2)x/k_{FS}}(v_\uparrow^*u_\downarrow\e{-\i\gamma_-^\uparrow}+u_\uparrow^*v_\downarrow\e{-\i\gamma_+^\downarrow+\i\Delta\theta}) + \mathcal{A}_4\e{-\i(2k_{FS}^2-k_y^2)x/k_{FS}}(u_\uparrow^*v_\downarrow\e{\i\gamma_-^\downarrow} + v_\uparrow^*u_\downarrow\e{\i\gamma_+^\uparrow+\i\Delta\theta})\Big\} ,
\end{align}
\end{widetext}
where we have defined $\xi_S = v_F/(|\Delta_\uparrow|+|\Delta_\downarrow|)$. The expression for the $y$-component of $\boldsymbol{\tau}_\text{STT}$ is obtained from Eq. (\ref{eq:tau}) by multiplication with an overall factor phase factor $\e{-\i\pi/2}$ inside the brackets $\Big\{\ldots\Big\}$. \addAB{The torque is perpendicular to the magnetization so its }
$z$-component 
vanishes\addAB{.} 
Both components of the spin transfer torque are proportional to the \addAB{injected} transverse 
spin current via the overall prefactor $cs$. The coefficients $\mathcal{A}_j$ depend on the coherence factors and wavevectors\addAB{,}
but are 
independent on the phase difference $\Delta\theta=\theta_\uparrow-\theta_\downarrow$ between the majority and minority spin superconducting order parameters.

Eq. (\ref{eq:tau}) 
demonstrates that the spin-transfer torque is qualitatively different in ferromagnetic superconductors as compared to ferromagnerts.
\addAB{The} torque 
has two terms proportional to $\e{\pm2\i hx/v_F}$ 
\addAB{that} correspond to the \addAB{conventional} 
rapid oscillations 
on a length scale $\lambda_h = 2\pi/(k_\uparrow-k_\downarrow) \sim 1/h$ which becomes of order $\mathcal{O}$(nm) in strong ferromagnets. 
However, there are two additional terms 
proportional to $\e{\pm\i(2k_{FS}^2-k_y^2)x/k_{FS}}$, which only appear in the presence of superconductivity ($\Delta_\sigma \neq 0$). Interestingly, these terms introduce a new and shorter length-scale 
due to the appearance of the term $\simeq 2k_F$ in the exponent (note that $k_{FS}\gg k_y$ due to the assumption $k_{FS}\gg k_{FN}$). The physical origin of these terms is 
particle-hole interference 
which \addAB{is unique in } 
the superconducting state \addAB{and vanishes } 
when $\Delta_\sigma\to 0$. 
\addAB{The injected spin-current causes the transmission of }
both electron-like and hole-like quasiparticles 
into the superconduct\addAB{or}.
The interference between two electron-like waves (or two hole-like waves) gives rise to the usual spin-transfer torque oscillating on the length scale $\lambda_h$. In contrast, the two last terms in Eq. (\ref{eq:tau}) \addAB{proportional to $u^*v$}  represent particle-hole interference. 
This also gives rise to a different length scale 
since 
hole-like waves have opposite momentum relative to their group velocity and thus interferes with the electron-like waves in a way that cancels the exchange-field dependence on the oscillation length. A unique aspect of the spin-transfer torque acting on a ferromagnetic superconductor is that the torque itself might be able to rotate the superconducting order parameter \cite{brataas_prl_04}. The latter, having a spin-triplet symmetry, is described by an orbital part and a vector in spin-space. For a sufficiently large torque acting on the magnetic order parameter, one might expect the superconducting order parameter to be rotated as well due to the coupling between them.

An 
intriguing feature about the spin-transfer torque in Eq. (\ref{eq:tau}) is that it depends explicitly on the difference $\Delta\theta$ between the spontaneously broken U(1) phases of the superconducting order parameters $\Delta_\sigma$. This is in contrast to \eg the charge-conductance, which is insensitive to $\Delta\theta$. This property of the spin-transfer torque may be understood as follows. For longitudinal spin currents, the
spin supercurrent is carried by the condensate with phase $\theta_\uparrow$ and the
condensate with phase $\theta_\downarrow$, but no superposition of these occurs. This
is different when a transverse spin current is injected 
\addAB{with a spin polarization at} an angle $\varphi$ with \addAB{respect to} the magnetic order parameter which corresponds to a non-collinear
superposition of quasiparticles from the two spin branches of the condensate. Therefore, the
phase difference appears in this contribution to the spin-transfer torque. As a result, the torque $\boldsymbol{\tau}_\text{STT}$ offers a possible probe for the relative phase difference $\Delta\theta$.

Let us investigate this in more detail. In the Hamiltonian used to model the coexistence of ferromagnetism and superconductivity, we have assumed that the spin-bands are independent by ignoring \eg spin-flip and spin-orbit scattering. Nevertheless, such processes can influence the relative superconducting phase between the bands due to a Josephson coupling between them. We can include these couplings by terms of the form $\lambda \text{Re}\{\Delta_\uparrow\Delta_\downarrow^\dag\}$ in the Ginzburg-Landau free energy. The coupling constant $\lambda$ depends on the system parameters and may change sign, which dictates whether the ground-state phase difference is $\Delta\theta=0$ (for $\lambda<0$) or $\Delta\theta=\pi$ (for $\lambda>0$). We have shown that the spin transfer torque depends on the phase difference $\Delta \theta$. Even in the scenario that there are only two possible values of the phase difference to $\Delta\theta\in \{0,\pi\}$, the absorbed torque is different in these two cases and the signatures of $\Delta \theta$ may thus be seen. We have verified this numerically (not shown) by considering the total torque absorbed after penetrating the ferromagnetic superconductor a distance $x$: $\boldsymbol{\tau}_\text{total} = \int^x_0 \sum_{k_y} \boldsymbol{\tau}_\text{STT}(x',k_y)\text{d}x'$, where $\boldsymbol{\tau}_\text{STT}$ is given by the general expression in Eq. (\ref{eq:termsc}). By including all transverse modes, classical dephasing has also been accounted for. We find that the torque is suppressed when $\Delta\theta=\pi$, which may be understood \addAB{from} 
Eq. (\ref{eq:tau}). Using a self-consistent calculation \cite{linder_prb_07}, 
$|\Delta_\uparrow|\simeq|\Delta_\downarrow|$ for a relatively weak exchange field $h\ll\mu_S$ as is relevant for experimentally observed ferromagnetic superconductors. As a result, $u_\sigma \simeq u_{-\sigma}$ and $v_\sigma \simeq v_{-\sigma}$. When $\Delta\theta=0$, the coherence factors in the terms $\e{\pm2\i hx/v_F}$, corresponding to the conventional spin-transfer torque, add constructively. However, a cancellation occurs when $\Delta\theta=\pi$, since $u_\sigma = v_\sigma^*$ for subgap energies (Fermi level). On the other hand, the torque terms due to particle-hole inteference in Eq. (\ref{eq:tau}) remain rather unchanged when changing $\Delta\theta$, as can be seen by using the abovementioned symmetries for the coherence factors.

\begin{figure}
\begin{center}
\includegraphics[width=9.cm]{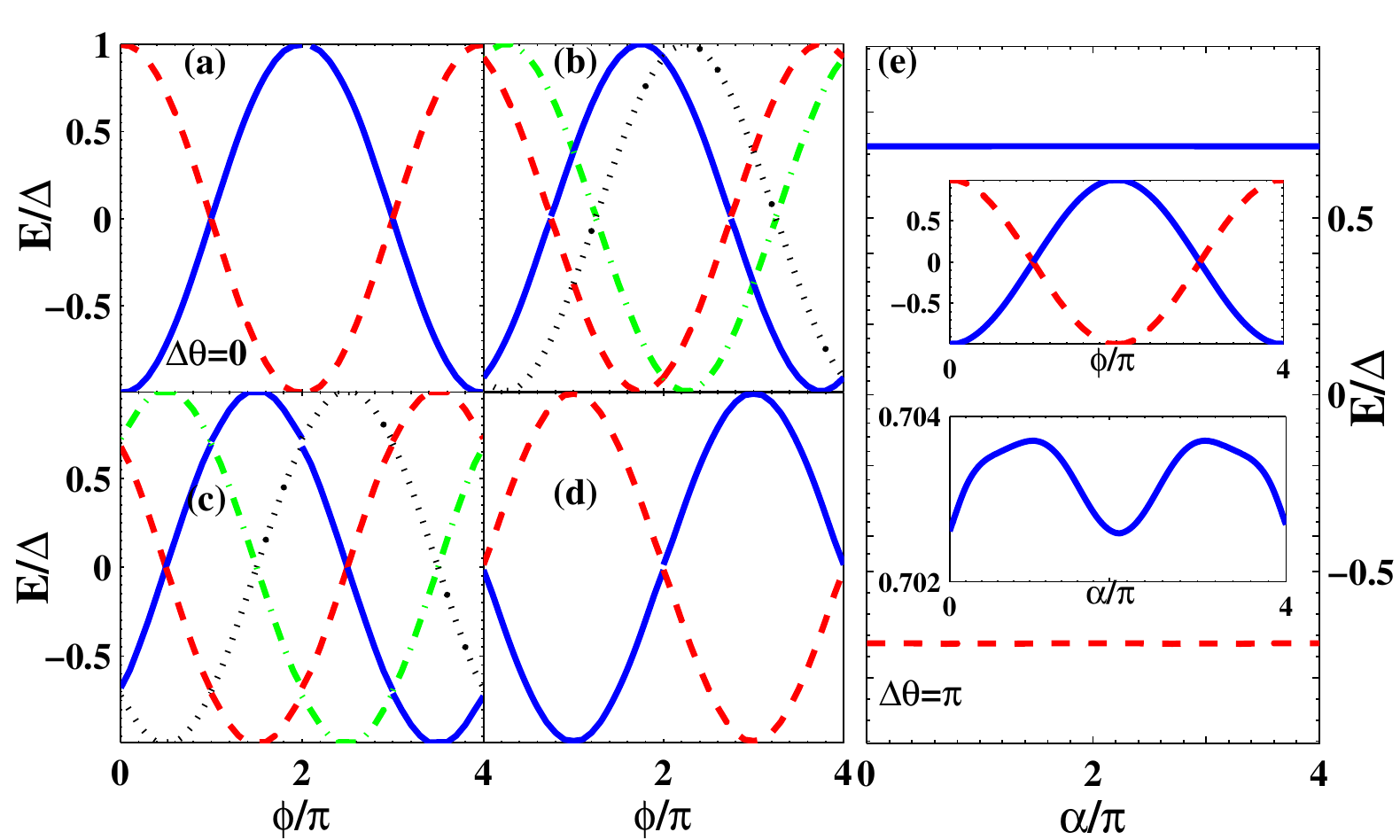}
\caption{ (color online)  Energy of Andreev bound states $E/\Delta$ for a short FS/N/FS contact consisting of four branches, two of which are degenerante for $\alpha=0$ (a) and $\alpha=\pi$ (d), (indicated in different colors) for the electron and hole excitations and for two directions of the spin. (a-d) For $\Delta \theta=0$ and versus the phase difference $\varphi$ for the angle $\alpha=0$ (a), $\pi/4$ (b), and $\pi/2$ (c), and $\pi$ (d). (e) For $\Delta \theta=\pi$ and when $\varphi=\pi/2$ as a function of $\alpha$ , 
where the energy is almost independent on the angle $\alpha$ (see also the lower inset) in contrast to the case of $\Delta \theta=0$.  The higher inset shows the dependence of $E/\Delta$ on $\varphi$ for $\alpha=\pi/2$. We have set $h/E_F=0.1$ and $h/\Delta=10$ and $k_y=0$. \label{fig1}}
\end{center}
\end{figure}

\begin{figure}
\begin{center}
\includegraphics[width=9.cm]{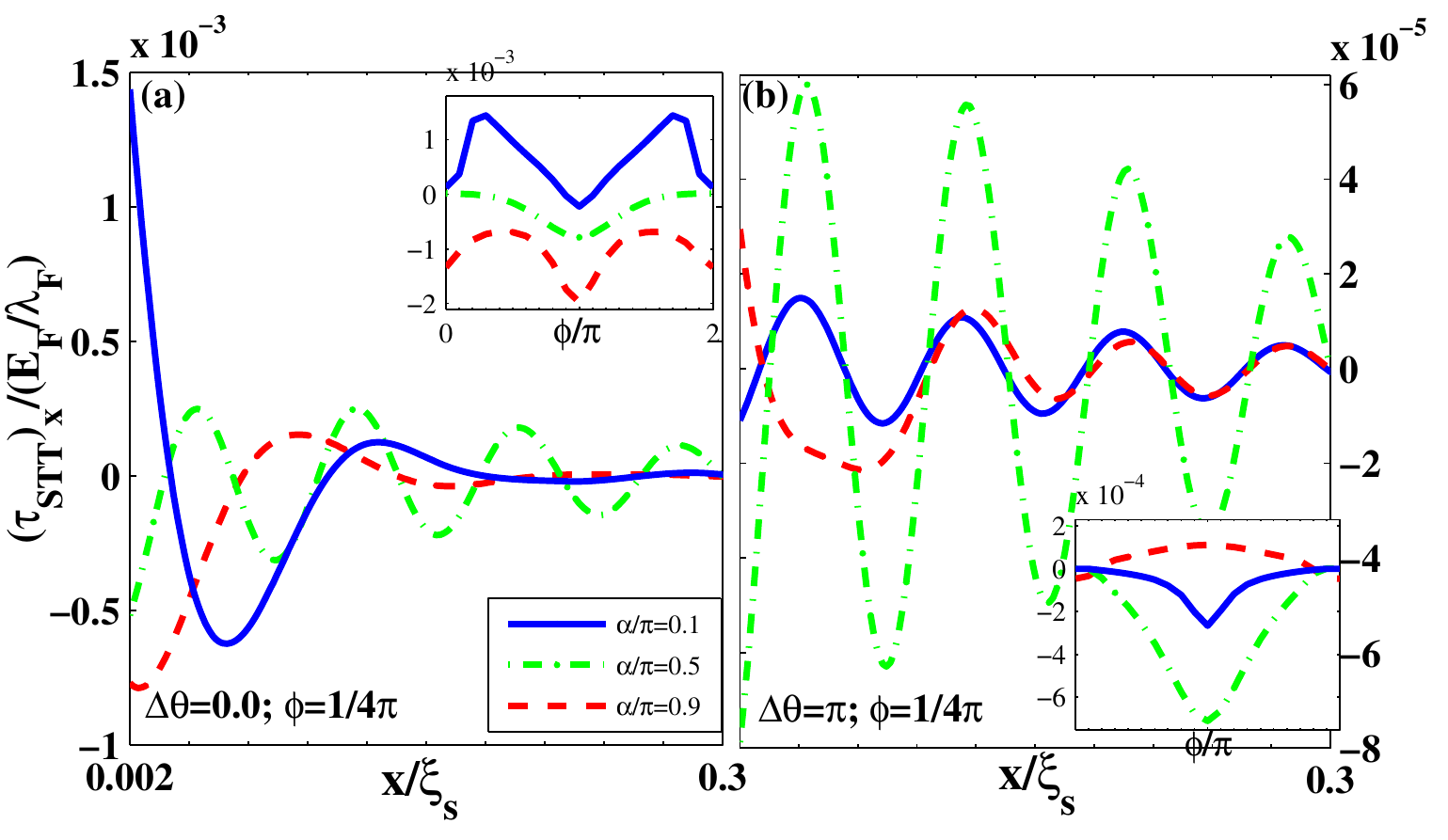}
\caption{ (color online) The component of the exchange torque that is perpendicular to the plane formed by  the magnetization vectors, $(\tau_\text{EXT})_x$, as a function of the distance $x$ from the FS/N interface inside FS1 when $\Delta \theta=0$ (a) and $\Delta \theta=\pi$ (b). We have set $\varphi=\pi/4$, $h/E_F=0.1$ and $h/\Delta=10$. The insets show the corresponding dependence on $\varphi$ calculated at the point $x=L=0.002 \xi$ for different angles $\alpha$.
\label{fig2}}
\end{center}
\end{figure}

We will now complement the understanding of the out-of-equilibrium spin-transfer torque with an equilibrium analogue that can be seen in the Josephson effect between 
ferromagnetic superconductors separated by a normal metallic (N) layer. In this case, in addition to the spin U(1) phase $\Delta \theta$ assumed to be identical in both FSs, there is also an overall phase difference $\varphi$ between the order parameters of the 
FSs. The magnetization vectors of \addAB{the} 
FSs are assumed to be misaligned by an angle $\alpha$. For subgap energies $|E|\leq |\Delta_{\uparrow,\downarrow}|$, successive AR at the 
N/FS interfaces and the coherent propagation of the excitations between these reflections lead to the formation of 
Andreev bound states \cite{LikhRev:79,Tinkham:04,Been:1991}. In our FS/N/FS system, these are 
correlated electron-hole pairs in spin-triplet states with a non-collinear polarization.
We have found the spectrum of these Andreev bound states by considering the electronic states in the 
the corresponding spinors and their derivatives at the right and the left N/FS-interfaces. We restrict ourselves to the 
case of a short N contact with thickness $L$ that is much smaller than the superconducting coherence length $\xi$. In this limit, the subgap Andreev states with $|E|\leq |\Delta_{\uparrow,\downarrow}|$
\addAB{dominates}
the superconducting transport properties 
\cite{BeenvanHout:1991}. The opening of a gap at the Fermi level removes part of the normal-state exchange torque, and we assume that the the Andreev-states dominate the net exchange interaction.

Fig. ~\ref{fig1} shows the dependence of the Andreev energy-states on the phase difference $\varphi$ for 
$\Delta \theta=0$ and for varying angles $\alpha$ [$\alpha=0$ (a), $\alpha=\pi/4$ (b), $\alpha=\pi/2$ (c) and $\alpha=\pi$ (d)] and $\Delta \theta=\pi$ (e) when $k_y=0$.   For both Figs. \ref{fig1} and \ref{fig2}, we use the same values for the exchange and superconducting gap, $h/E_F=0.1$, $h/\Delta=10$ and assumed transparent interfaces. 
The Andreev bound states consist of four branches in the space of electron-hole excitations and in the space of spin.  Fig.~\ref{fig1} (a-d) show that for $\Delta \theta=0$ the phase dependence for these branches is shifted by $\pm \alpha$ with respect to that of a conventional short SNS junction, which is given by $\epsilon_{e,h}=\pm \Delta_0\cos{(\varphi/2)}$\cite{KulicOmelyn:7778}. Thus, considering transport normal to the interface, the Andreev energies closely obey the relations $\epsilon_{e,h,\uparrow,\downarrow}=\pm \Delta_0\cos{[(\varphi\pm\alpha)/2]}$ 
when $\Delta \theta=0$. Note that for the collinear configurations ($\alpha = 0$ or $\pi$) the branches are doubly degenerate.  In contrast, when $\Delta \theta=\pi$, the dependence on $\alpha$ is rather weak, as is seen in Fig. \ref{fig1}e (see also the lower inset). In this case the phase dependence is almost the same as in the conventional SNS case, as seen in the higher inset of Fig. \ref{fig1}e.

The spin-polarized Andreev states can carry charge and spin supercurrent. The spin supercurrent is related to the magnetic coupling originating from superconducting correlations between the magnetization vectors of the two FSs. As a result, an equilibrium exchange torque is exerted on the magnetization vectors. We have calculated the $x$-component of this Andreev torque, $(\tau_\text{EXT})_x$, using Eq. (\ref{eq:termsc}) and by summing over the contribution of all Andreev states. This component is perpendicular to the plane formed by the two magnetization vectors and tends to rotate them around the $x$-axis. We have found that this superconducting torque is odd in $\alpha$ but even in $\varphi$, obeying the relations $(\tau_\text{EXT})_x(\alpha,\varphi)=-(\tau_\text{EXT})_x(2\pi-\alpha,\varphi)$ and  $(\tau_\text{EXT})_x(\alpha,\varphi)=(\tau_\text{EXT})_x(\alpha,2\pi-\varphi)$, respectively.

Fig.~\ref{fig2} presents the dependence of $(\tau_\text{EXT})_x$, acting on the magnetization in FS1, on the distance $x$ from the N/FS interface and for different values of $\alpha$ when $\Delta \theta=0$ (a) and $\Delta \theta=\pi$ (b).   We have here fixed $\varphi=\pi/4$. 
$(\tau_\text{EXT})_x$ \addAB{exhibits} 
spatial oscillations with a period that  \addAB{around} 
$\lambda_h\sim 1/h$ 
\addAB{and the} 
amplitude of  
decay\addAB{s} with $x$ over a penetration length which is a fraction of $\xi$. For $\Delta \theta=\pi$, the amplitude of the exchange torque is diminished, as compared to $\Delta \theta=0$ case. This behavior is qualitatively the same as for the non-equilibrium spin-transfer torque.
The dependence on the phase $\varphi$ is shown in the insets of Fig. ~\ref{fig2}a,b. As can be seen, for both values of $\Delta \theta$ the amplitude as well as the direction of the torque can be tuned by externally changing the phase difference over the contact.

In conclusion, 
in ferromagnetic superconductors, \addAB{there is a novel spin-transfer torque} which arises from particle-hole interference between quasiparticles from the two spin branches of the condensate. In a normal metallic contact between two ferromagnetic superconductors, an equilibrium spin Josephson current arises, which results from non-collinear magnetizations and is carried by spin-triplet Andreev bound states \addAB{and is sensitive to the spin-resolved (internal) phase difference and the applied phase difference between the superconcucting order parameters}. 
These findings could open new perspectives for obtaining phase-dependent spin-polarized transport and magnetization dynamics by combining ferromagnetic and superconducting correlations.

\textit{Acknowledgments.} J. L. would like to thank A. Sudb{\o} for many useful discussions. Z. S. and M. Z. 
acknowledge support by the Institute for Advanced Studies in Basic Sciences (IASBS) Research Council under grant No. G2012IASBS110. M. Z. thanks ICTP in Trieste for hospitality and support\addAB{.}

\end{document}